\documentclass[aps, prd, superscriptaddress, twocolumn,a4paper]{revtex4-1}

\usepackage{hyperref}
\usepackage{graphicx}
\usepackage{amsmath}
\usepackage{color}
\usepackage{float}
\usepackage{bm}
\usepackage[english]{babel}

\newcommand{\be}{\begin{equation}}
\newcommand{\ee}{\end{equation}}
\newcommand{\ba}{\begin{eqnarray}}
\newcommand{\ea}{\end{eqnarray}}
\newcommand{\nn}{\nonumber\\}
\newcommand{\KMAX}{k_\mathrm{max}}
\begin{document}
\title{Energy Loss Versus Energy Gain of Heavy Quarks in a Hot Medium}

\author{Mohammad Yousuf Jamal}
\affiliation{School of Physical Sciences, National Institute of Science Education and Research, HBNI, Jatni-752050, India}

\author{Santosh K. Das}
\affiliation{School of Physical Sciences, Indian Institute of Technology Goa, Ponda-403401, Goa, India}

\author{Marco Ruggieri}
\affiliation{School  of  Nuclear  Science  and  Technology,  Lanzhou  University,
222  South  Tianshui  Road,  Lanzhou  730000,  China}

\begin{abstract}
We study the energy loss and the energy gain of heavy quarks in a hot thermal medium. These include the study of the energy change due to the polarization and to the interaction with the thermal fluctuations of the medium. The dynamics of the heavy quarks with the medium is described by the Wong equations, that allow for the inclusion of both the backreaction on the heavy quarks due to the polarization of the medium, and of the interaction with the thermal fluctuations of the gluon field. Both the momentum as well as the temperature dependence of the energy loss and gain of charm and bottom quark are studied. We find that heavy quark energy gain dominate the energy loss at high-temperature domain achievable at the early stage of the high energy collisions. This finding supports the recently observed heavy quarks results in Glasma and will have a significant impact on heavy quark observables at RHIC and LHC energies.   
\end{abstract}

\keywords{Relativistic heavy ion collisions, Glasma, Yang-Mills fields, heavy quarks, 
energy change, Debye mass, Quark- Gluon plasma}

\maketitle
\section{Introduction}
The medium consisting of quarks and gluons produced at various heavy-ion collision experimental facilities such as 
the Relativistic Heavy Ion Collider (RHIC) and the Large Hadron Collider (LHC)   
provide a unique opportunity to explore the Quantum Chromodynamics (QCD) matter under extreme conditions of temperature and density. 
The bulk properties of such a state of matter, called Quark-Gluon Plasma (QGP)~\cite{Shuryak:2004cy, Science_Muller}, 
are governed by the light quarks and gluons. 
Though the small size and short-lived nature of the produced medium do not allow to observe it by the naked eyes and hence, 
we rely on the signatures observed at the detector end in the form of particle spectra. 

Heavy quarks, namely Charm and Beauty,, are considered as excellent 
probes of the QGP~\cite{Prino:2016cni,Andronic:2015wma,Rapp:2018qla,Aarts:2016hap,Dong:2019unq,Cao:2018ews,Xu:2018gux,Scardina:2017ipo} 
and offers signatures of the production of the QGP itself.
In fact, one of these signatures 
is the suppression of high $p_T$ heavy hadrons~\cite{phenixelat,stare,alice}, that is understood as a result of the loss of energy
of the high-energy charm and beauty quarks while they propagate through the dense matter formed after collisions. 
More generally, 
the energy change of charm and beauty in the QGP have two major contributions, namely 
the polarization of the medium, which leads to energy loss, and the interaction with the background thermal fluctuations
fluctuations of the gluon field that is responsible of momentum diffusion. 
The polarization is responsible of energy loss~\cite{Han:2017nfz, Jamal:2019svc} while interaction with
thermal fluctuations leads to energy gain and is effective in the low-velocity limit~\cite{Chakraborty:2006db,Sheikh:2019xjs},
see also~\cite{bjorken1982energy,Thoma:1990fm,Braaten:1991jj,Mrowczynski:1991da, Thomas:1991ea, Koike:1992xs,
Romatschke:2004au, Baier:2008js, Carrington:2015xca,Jiang:2014oxa, Jiang:2016duz,
Baier:2000mf, Jacobs:2004qv, Armesto:2011ht, Majumder:2010qh, Mustafa:1997pm, Dokshitzer:2001zm, 
	Djordjevic:2003zk, Wicks:2007am, Abir:2011jb, Qin:2007rn, Cao:2013ita, Mustafa:2003vh, DuttMazumder:2004xk, 
	Meistrenko:2012ju, Burke:2013yra, Peigne:2007sd, Neufeld:2014yaa, Chakraborty:2006db, Adil:2006ei, 
	Peigne:2005rk, Dusling:2009jn, Cho:2009ze,Elias:2014hua,Han:2017nfz,Fadafan:2008gb, Fadafan:2008uv, Fadafan:2012qu, 	Abbasi:2012qz, Abbasi:2013mwa}.
Heavy quarks can experience diffusion in the early stage of high energy nuclear collisions as well.
In particular, recent studies suggest that due to the high energy density developed in the early stage,
the motion of charm and beauty is dominated by field fluctuations that lead to a modest energy gain
of the heavy probes and to a tilt in the spectrum~\cite{Ruggieri:2018rzi, Sun:2019fud, Liu:2019lac,Mrowczynski:2017kso,Ruggieri:2019zos},
in qualitative agreement with previous studies on the propagation in a high temperature
QGP medium~\cite{Chakraborty:2006db, Sheikh:2019xjs}.

The purpose of the present study is to analyze the combined effect of energy gain and energy loss of heavy quarks
in a high temperature QCD medium, analyzing the kinematic regime in which one of the two mechanisms dominates.
In solving this problem, albeit using several approximations, we will show that even when energy gain and
energy loss are considered consistently, the heavy quarks will experience a substantial energy gain if the
temperature of the medium is large enough. 
We will address quantitatively the question which between energy gain or energy loss of heavy quarks is dominant
in a given kinematic regime and at a given temperature. 
The conclusion is easy to imagine: if the temperature is quite larger than the kinetic energy of the heavy quark,
then the medium will contribute substantially to increase the energy of the heavy probe as this propagates in the
hot medium; energy loss will be important when the temperature of the medium is lower than the kinetic energy 
of the heavy quark.  
These qualitative statements need to be supported by quantitative findings, aiming to identify the kinematic regimes
in which energy loss or energy gain dominate and thus giving a clearer understanding of the dynamics of heavy quarks in the
QGP medium produced in collisions. This is what we want to study here.

In addition to this, our results 
offer a case study that supports the assumption of 
 \cite{Liu:2019lac,Ruggieri:2018rzi,Sun:2019fud, Carrington:2020sww} where heavy quarks propagate in the evolving Glasma fields,
and in which energy loss has been ignored. 
In fact, although here we consider a thermalized medium while the evolving Glasma is out of equilibrium,
the diffusion of heavy probes in Glasma resembles that in a thermal medium, at least when an average over the 
heavy quark spectrum is considered (see for example Fig. 7 of~\cite{Liu:2019lac});
the energy density in the evolving Glasma is very large, implying that the effective temperature of the medium is also
high and thus the loss of energy of low momentum quarks can be neglected.

The plan of the paper is as follows. In section~\ref{sec:rl}, we shall discuss the polarization energy loss of heavy 
quarks moving in the hot QCD medium along with a brief description of the change in energy of heavy quarks due to fluctuation.
In section~\ref{sec:res}, we shall discuss the various results.
Section~\ref{el:saf}, is dedicated to the summary and future possibilities of the present work.
	
\section{Energy change due to polarization and fluctuation }
\label{sec:rl}
In this section we discuss the theoretical setup on which we base our analysis.
We treat charm and beauty 
as classical color sources that obey the Wong's equations~\cite{Wong:1970fu};  
these equations describe the motion of classical colored particles interacting 
with a dynamical gluon field, $F_{a}^{\mu\nu}$, and in a Lorentz covariant form they are given by
\ba
\frac{dx^{\mu}(\tau)}{d\tau} &=& u^{\mu}(\tau),\label{eq:1_1} \\
\frac{dp^{\mu}(\tau)}{d\tau} &=&g q^{a}(\tau)F^{\mu\nu}_{a}(x(\tau))u_{\nu}(\tau),\label{eq:1_2} \\
\frac{dq^{a}(\tau)}{d\tau} &=& -gf^{abc}u_{\mu}(\tau)A^{\mu}_{b}(x(\tau))q_{c}(\tau);
\label{eq:wong}
\ea 
in these equations, $q^a(\tau)$ is a classical charge (to not be confused with the fundamental,
quantized color charge of the quark) that is introduced to describe the conservation of the color current
in the classical theory, with $a-1,2,\dots,N_c^2-1$, $g$ is the coupling constant, 
$\tau$,  $x^{\mu}\equiv X $, $u^{\mu}=\gamma(1,{\bm v})$ and $p^{\mu}(\tau)$ are the proper time, trajectory, 
4-velocity  and 4-momentum of the heavy quark, respectively. 
For $N_c$ fundamental colors of quarks there are $\text N_{c}^{2}-1$ chromo-electric/magnetic fields, 
and $f^{abc}$ is the structure constant  of ${\text {SU}({\text N_c})}$ gauge group; finally, $A^{\mu}_{a}$ is the gauge potential.
In solving these equations we assume the gauge condition $u_{\mu}A^{\mu}_{a}(X) = 0$ \cite{Jiang:2016duz, Carrington:2015xca},
namely that the gauge potential vanishes on the trajectory of the particle and 
which implies that $q^{a}$ is independent of $\tau$; moreover, we assume that 
in the motion of the heavy quark in the thermal medium the magnitude of the velocity does not change much \cite{Jiang:2016duz}.

From the $\mu=0$ component of Eq.~\eqref{eq:1_2} 
the energy change per unit time is
\ba
\frac{dE}{dt}=g~q^a~\bm v\cdot {\bm E}^{a}(X),
\label{eq:el1}
\ea
where here and in the following we use $E$ to denote the energy of the heavy quark and $\bm E$ for the color-electric field,
and $t=\gamma\tau$ is the time in the laboratory frame in which the heavy quark of mass $M$ moves with velocity $\bm v =\frac{{\bm p}}{\sqrt{{ p}^2+M^2}}$.
The color field consists of two terms,
\begin{equation}
{\bm E}^{a} = {\bm E}_\mathrm{ind}^{a} + {\bm E}_\mathrm{fluct}^{a},\label{eq:indPLUSfluc}
\end{equation}
where ${\bm E}_\mathrm{ind}^{a}$ denotes the field induced by the motion of the heavy quark that polarizes medium
(for this reason, this is also called the polarization contribution),
hence representing an energy loss and its inclusion in the equation of motion
amounts to consider the backreaction on the heavy quark, while  ${\bm E}_\mathrm{fluct}^{a}$ denotes the color field
induced by the thermal fluctuations in the gluon medium: the interaction of the heavy quark
with ${\bm E}_\mathrm{fluct}^{a}$ can result in energy loss or energy gain depending on the temperature
of the medium as well as on the heavy quark momentum, as we discuss later.

For the motion of the heavy quark in a thermal medium, the right hand side of  Eq.~\eqref{eq:el1} is replaced by its ensemble average, 
\ba
\frac{dE}{dt}=g~q^a~\langle {\bm v}(t)\cdot {\bm E}^{a}(X(t))\rangle,
\label{eq:el2}
\ea
where the electric field is given by Eq.~\eqref{eq:indPLUSfluc}.
The procedure to evaluate right hand side of the above equation is explained clearly in the literature, see for 
example~\cite{Chakraborty:2006db},
therefore we limit ourselves to quote the final result that is
\ba
\frac{dE}{dt}&=&\left< g~q^a {\bm v_0} \cdot {\bm E^a}\right > \nn 
&+& g^2~\frac{q^a q^b}{E_0} \int_{0}^{t}dt_1\left<{\bm E^b_t (t_1)}\cdot {\bm E^a_t(t)}\right > \nn 
&+& g^2~\frac{q^a q^b}{E_0} \int_{0}^{t}dt_1\int_{0}^{t}dt_2\bigg< \Sigma_j{\bm E^b_{t,j} (t_2)}\nn
&\times & \frac{\partial}{\partial {\bm r_{0j}}} {\bm v_0}\cdot {\bm E^a_t(t)}\bigg >.
\label{el_full}
\ea
Equation~(\ref{el_full}) corresponds to the full energy change of the heavy quark: 
the first addendum on the right hand side is the energy loss due to the work against the induced field that has been
discussed in the previous subsection, while the remaining addenda correspond to the change of energy 
due to the thermal fluctuations of the gluon fields. 
In the intermediate steps it has been assumed that $\langle {\bm E^a_i}{\bm B^a_j} \rangle=0$ and $\langle {\tilde{{\bm E}}}\rangle=0$.
We rewrite Eq.~\eqref{el_full} as
\begin{equation}
\frac{dE}{dt} = \left(\frac{dE}{dt}\right)_\mathrm{ind} + \left(\frac{dE}{dt}\right)_\mathrm{fluct},
\end{equation}
where
\begin{equation}
\left(\frac{dE}{dt}\right)_\mathrm{ind} = \langle g~q^a {\bm v_0}\cdot {\bm E^a}\rangle \label{eqLell}
\end{equation}
and
\begin{eqnarray}
\left(\frac{dE}{dt}\right)_\mathrm{fluct} &=& g^2~\frac{q^a q^b}{E_0} \int_{0}^{t}dt_1\left<{\bm E^b_t (t_1)}\cdot {\bm E^a_t(t)}\right > \nn 
&+& g^2~\frac{q^a q^b}{E_0} \int_{0}^{t}dt_1\int_{0}^{t}dt_2\bigg< \Sigma_j {\bm E^b_{t,j} (t_2)}\nn
&\times & \frac{\partial}{\partial {\bm r_{0j}}} {\bm v_0}\cdot {\bm E^a_t(t)}\bigg >,\label{eqDDD}
\end{eqnarray}
and we discuss the two terms separately below.

\subsection{Energy loss due to the induced field}
Firstly we analyze the energy loss due to the work against the induced field,
see Eq.~\eqref{eqLell} \cite{Thoma:1990fm, Koike:1992xs, Koike:1991mf, Han:2017nfz}.
The induced field can be obtained by solving the Yang-Mills equations for a thermalized gluon system with the source given by
the color current carried by the heavy quark, namely~\cite{Jamal:2019svc},
\ba
{\bm E}^{a}_\mathrm{ind}(X) &=&-i\frac{gq^{a}}{\pi}
\int \text{d}\omega \text{d}^3k\frac{1}{\omega~ k^2}\bigg[{\bm k}~({\bm k}\cdot{\bm v})\left(\epsilon^{-1}_L-1\right)
\nn &+& \Big({k}^2 {\bm v}-{\bm k}~({\bm k}\cdot{\bm v})\Big) \bigg\{\left(\epsilon_T-\frac{k^2}{\omega^{2}}\right)^{-1}
\nn &-&\left(1-\frac{k^2}{\omega^2}\right)^{-1}\bigg\}\bigg]\frac{e^{ i({\bm k}\cdot{\bm x}-\omega t)}}{\omega -{\bm k}\cdot{\bm v}+i 0^+};
\label{eq:inelc}
\ea
performing the $\omega$ integration in Eq.~\eqref{eq:inelc} and subtituting in Eq.~\eqref{eqLell} we get
	\ba
\left(\frac{dE}{dt}\right)_\mathrm{ind} &=&-\frac{C_F \alpha _s}{2 \pi ^2 |{\bm v}|}\nonumber\\
&&\times
\int^{k_{max}}_{k_0} d^3{\bm k}\frac{\omega }{k^2}\bigg\{\left(k^2 |{\bm v}|^2-\omega ^2\right)
\text{Im}\frac{1}{\omega ^2 \epsilon_T-k^2}\nn
&+&\text{Im}\frac{1}{\epsilon_{L}}\bigg\}_{\omega =\bm k \cdot \bm v},
\label{eq:de}
\ea
where, $k^\mu= (\omega, {\bm k})$ with $|{\bm k}|=k$ and  $\alpha_{s}$ is the QCD coupling;
moreover,
$C_F =4/3$ is the Casimir invariant in the fundamental representation of the $\text {SU}({\text N_c})$, 
$\epsilon_{L}$ and $\epsilon_{T}$ are the longitudinal and transverse components of the medium dielectric permittivity,
that have been computed using the semi-classical transport theory approach in~\cite{Jamal:2019svc}.

\subsection{Interaction with the fluctuating field}

The energy change due to the interaction of the heavy quark with the fluctuating field  can be written as~\cite{Chakraborty:2006db},
	\ba
\left(\frac{dE}{dt}\right)_\mathrm{fluct}&=&\frac{C_F \alpha _s}{8 \pi ^2 E_0 v^4}
\int^{\KMAX v}_{0} d\omega \coth\frac{\beta \omega }{2} F(\omega,k=\omega/v)\nn &+&
\frac{C_F \alpha _s}{8 \pi ^2 E_0 v^2}
\int^{\KMAX}_{0} dk k \int^{k}_{0}d\omega \coth\frac{\beta \omega }{2} G(\omega,k),\nn
&&
\label{eq:elfl}
\ea 
where
\ba
F(\omega,k)&=& 8 \pi \omega^2 \frac{\text {Im}[\epsilon_L]}{|\epsilon_L|^2},\nn
G(\omega,k)&=& 16 \pi \frac{\text {Im}[\epsilon_T]}{|\epsilon_T-k^2/\omega^2|^2}.
\label{eq:fl}
\ea
In Eq.~\eqref{eq:elfl} we have put $E_0=\sqrt{p^2+M^2} $
and introduced an ultraviolet cutoff, $\KMAX$, 
which is of the order of the Debye screening mass \cite{Chakraborty:2006db, Adil:2006ei};
in the following we will consider two representative values of this cutoff,
namely $\KMAX=m_D$ and $\KMAX=2m_D$: while the specific value of 
$\KMAX$ affects the results quantitatively, the qualitative picture is almost unaffected by this choice.

 \section{Results}
\label{sec:res}

In this section we summarize our results: firstly we focus on charm, then we turn on beauty. 
We use the set of parameters $N_c=3$, $N_f=2$ and  $\alpha_s=0.3$.
In all the figures below we show the energy change per unit length since the latter is the most used in the literature: 
this can be obtained easily
from the change of energy per unit time that we have computed in the previous section,
\begin{equation}
\frac{dE}{dx} = \frac{1}{|\bm v|}\frac{dE}{dt},
\end{equation}
where $\bm v$ is the velocity of the heavy quark. Moreover, to uniform to the existing literature
we plot $-dE/dx$ since this quantity is been mostly used to quantify the energy loss and is therefore positive.

\subsection{Charm}

\begin{figure}[t!]
	\centering
	\includegraphics[height=5cm,width=6.0cm]{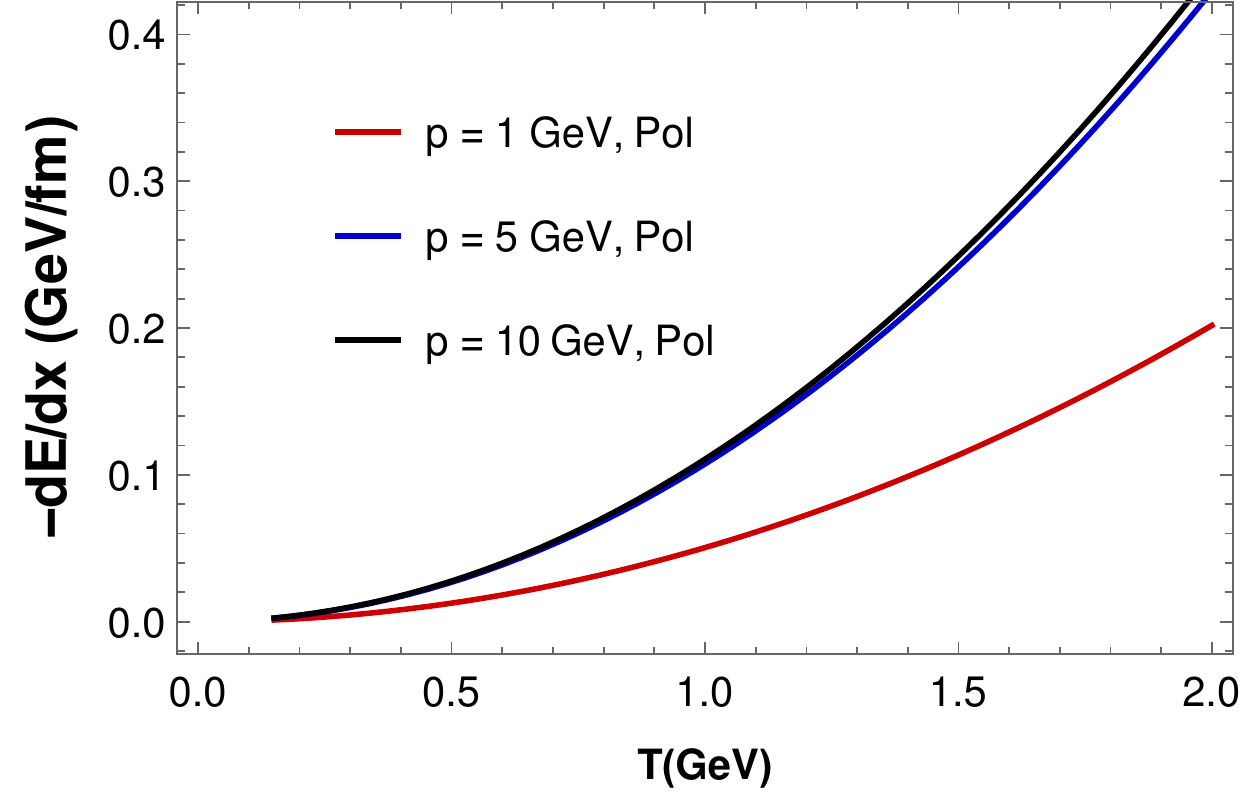}\\
	\includegraphics[height=5cm,width=6.0cm]{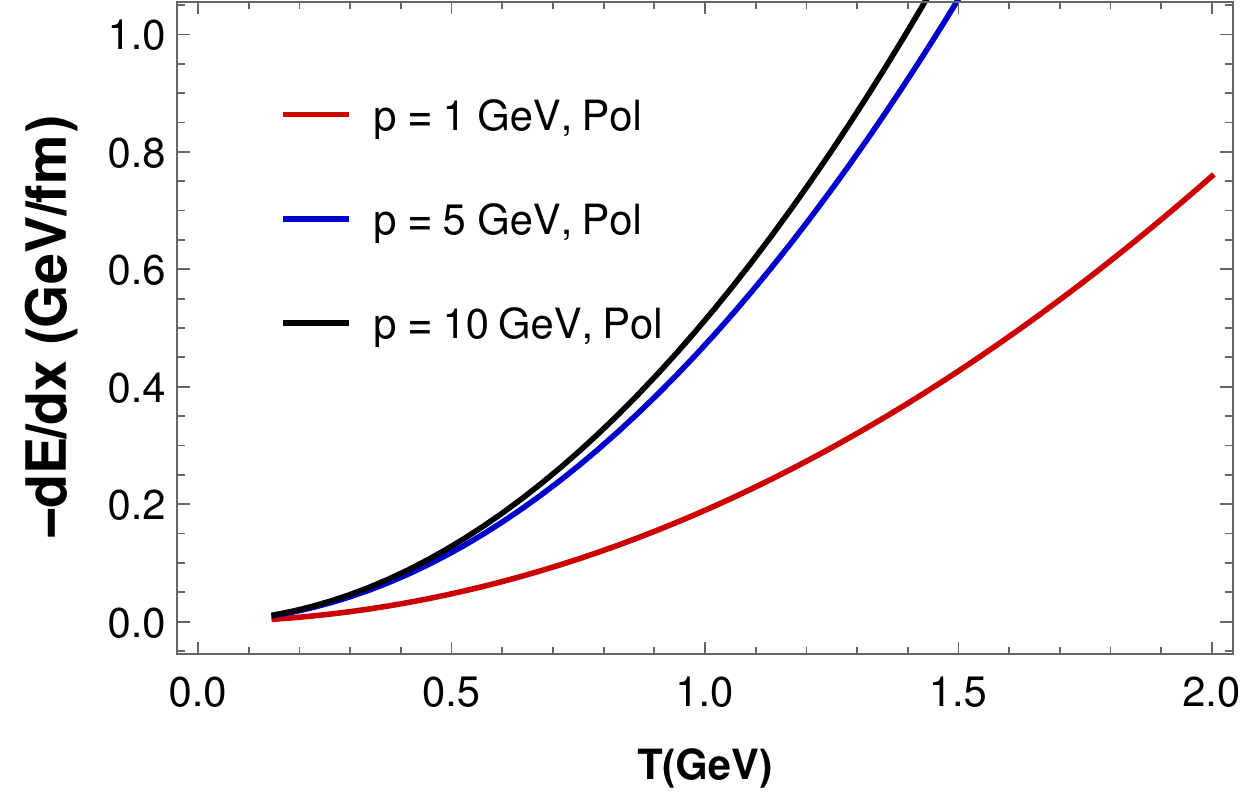}
	\caption{Energy loss of charm due to the polarization of the hot medium, $-(dE/dx)_\mathrm{ind}$,
	versus temperature, for three values of the initial charm quark momentum. 
	Upper and lower panels correspond to $\KMAX=m_D$ and $\KMAX=2m_D$ respectively.}
	\label{fig:pol_charm}
\end{figure}

In Fig.~\ref{fig:pol_charm} we plot $-(dE/dx)_\mathrm{ind}$ versus temperature for three values of the heavy quark momentum.
In the figure, upper and lower panels correspond to $\KMAX=m_D$ and $\KMAX=2m_D$ respectively.
As anticipated, the backreaction represented by the interaction of the heavy quark with the induced field
results in an energy loss. This can be understood easily since the motion of the heavy quark in the thermal medium
results in the polarization of the medium itself, and for this process to happen energy has to be transferred from the quark
to the medium itself. For example, for a charm quark with initial momentum $p=10$ GeV, at a temperature $T= 1$ GeV
we find $-(dE/dx)_\mathrm{ind}\approx 0.1$ GeV/fm for $\KMAX=m_D$ and
$-(dE/dx)_\mathrm{ind}\approx 0.5$ GeV/fm for $\KMAX=2m_D$.

\begin{figure}[t!]
	\centering
	\includegraphics[height=5cm,width=6.0cm]{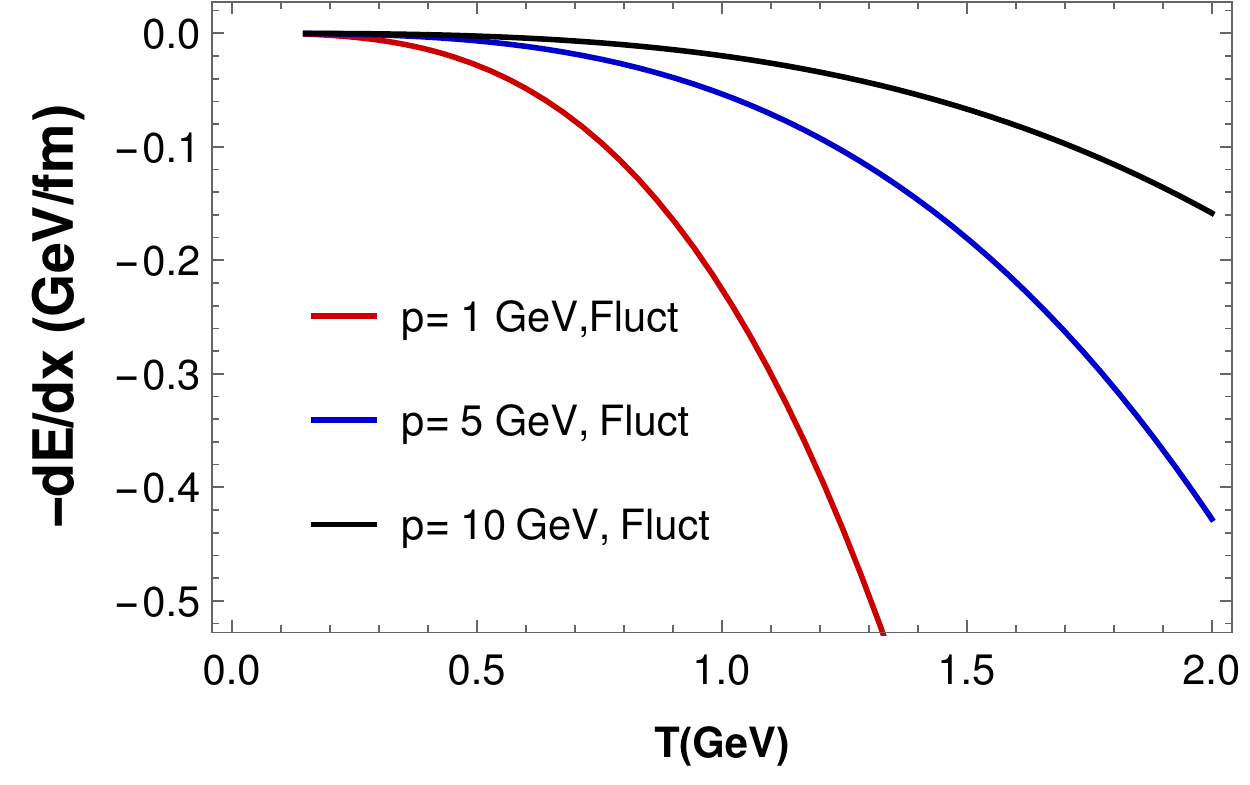}\\
	\includegraphics[height=5cm,width=6.0cm]{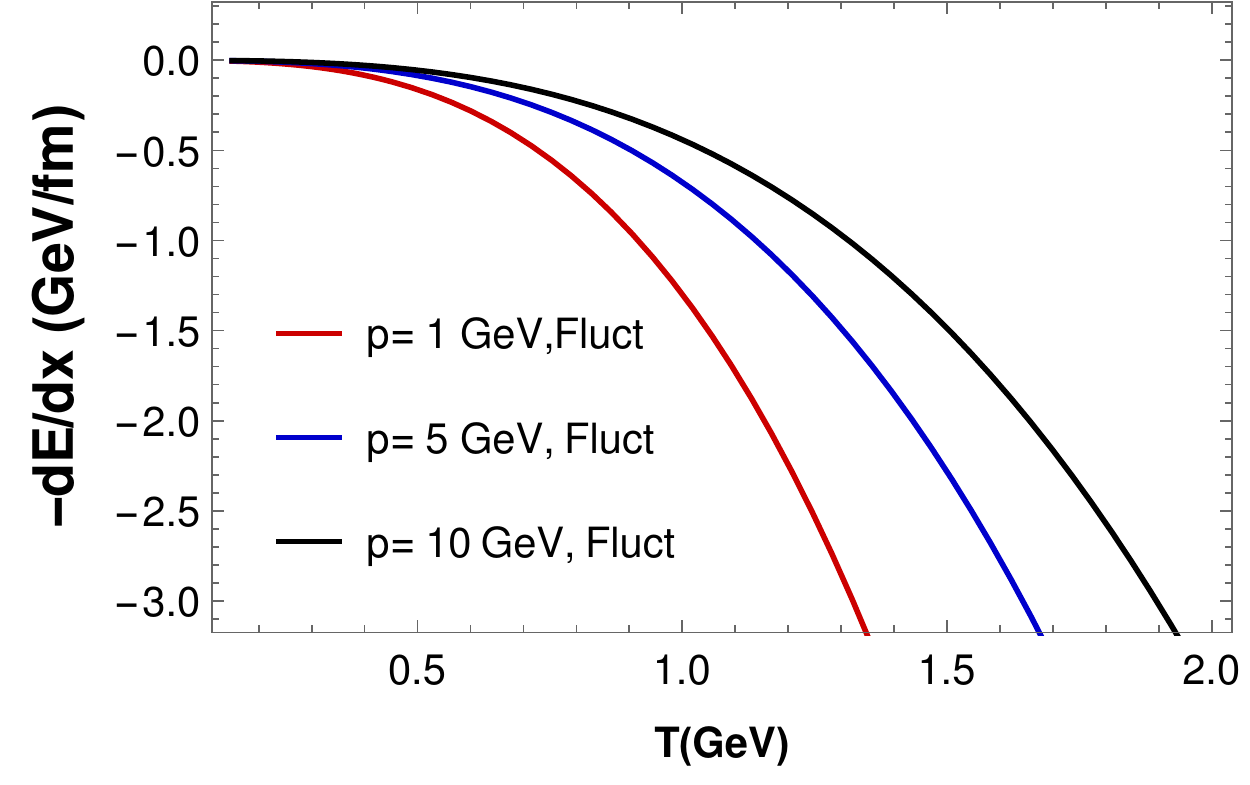}
	\caption{Energy change of charm due to the fluctuation of the hot medium, $-(dE/dx)_\mathrm{fluct}$,
	versus temperature, for three values of the initial charm quark momentum. 
	Upper and lower panels correspond to $\KMAX=m_D$ and $\KMAX=2m_D$ respectively.}
	\label{fig:fl_charm}
\end{figure}

In Fig.~\ref{fig:fl_charm} we plot $-(dE/dx)_\mathrm{fluct}$ versus temperature for three values of the initial heavy quark momentum;
upper and lower panels correspond to $\KMAX=m_D$ and $\KMAX=2m_D$ respectively.
Differently from the cases shown in Fig.~\ref{fig:pol_charm}, we find that the interaction with the thermalized gluon field
leads to energy gain rather than energy loss. For example, considering again $p=10$ GeV and $T=1$ GeV
we find $-(dE/dx)_\mathrm{fluct}\approx -0.02$ GeV/fm for $\KMAX=m_D$ and
$-(dE/dx)_\mathrm{fluct}\approx -0.4$ GeV/fm for $\KMAX=2m_D$.
The results shown in Figg.~\ref{fig:pol_charm} and~\ref{fig:fl_charm} agree qualitatively with those obtained within 
a purely classical model for the diffusion and the energy loss in a Brownian motion \cite{Ruggieri:2019zos},
in which the backreaction as the source of the energy loss and the interaction with the thermal fluctuations
as resulting in energy gain and momentum broadening appear clearly.
In addition to this, comparing the results shown in Figg.~\ref{fig:pol_charm} and~\ref{fig:fl_charm} we notice that
for $p/T\gg 1$ the energy loss due to polarization of the medium is larger than the energy gain,
but this situation changes when $p/T\lesssim 1$.

\begin{figure}[t!]
	\centering
	\includegraphics[height=5cm,width=6.0cm]{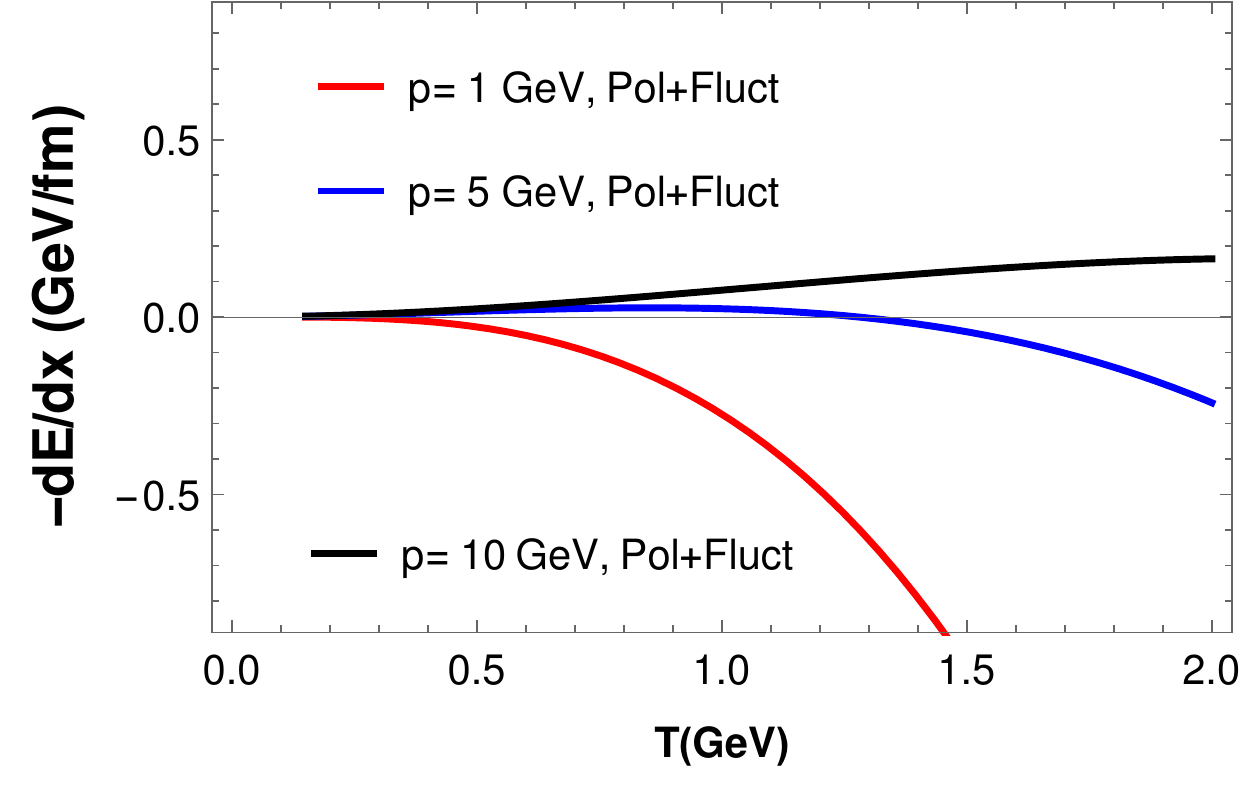}\\
	\includegraphics[height=5cm,width=6.0cm]{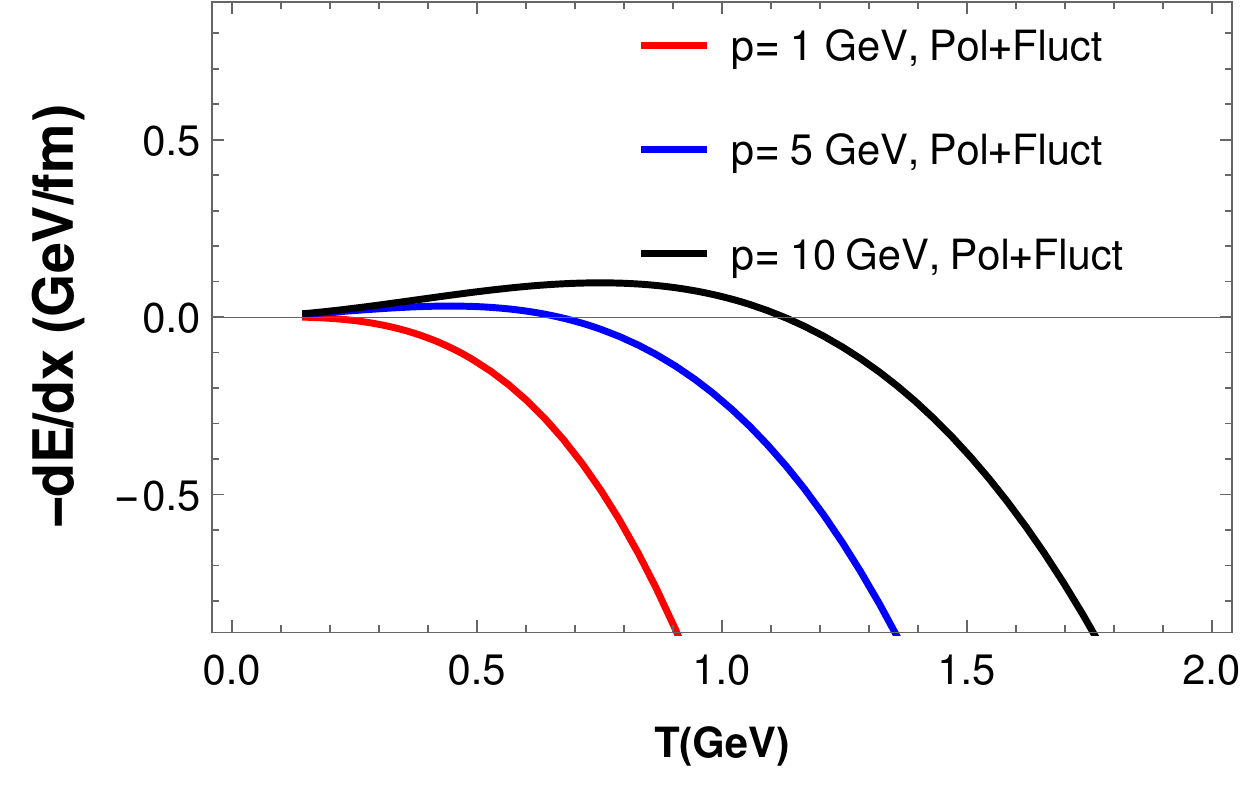}
	\caption{Energy change of charm quark due to fluctuation and polarization for $\KMAX= m_D$ (top) and $\KMAX= 2m_D$ (bottom). }
	\label{fig:full}
\end{figure}

In Fig.~\ref{fig:full} we plot the total energy change per unit length of the charm quark versus temperature,
for three values of the initial momentum $p$; this is obtained by adding the results shown in Figg.~\ref{fig:fl_charm} 
and~\ref{fig:pol_charm} . 
In Fig.~\ref{fig:full} the upper and lower panels correspond to $\KMAX=m_D$ and $\KMAX=2m_D$ respectively.
We notice that for $p=1$ GeV,
in the full range of temperature considered the sum of the polarization and  the fluctuation contributions
results in an energy gain of the quark. For the other two representative values of $p$,
namely for $p=5$ GeV and $p=10$ GeV, we find that up to $T\approx 1$ GeV
the charm losses energy by polarization of the medium, while for higher temperatures
it gains energy from the medium itself; the exception that we find is that if the initial momentum is very large,
see $p=10$ GeV in the figure, and $\KMAX=m_D$ then energy loss dominates over energy gain
over the whole range of temperature studied.
 
\begin{figure}[t!]
	\centering
	\includegraphics[height=5cm,width=6.0cm]{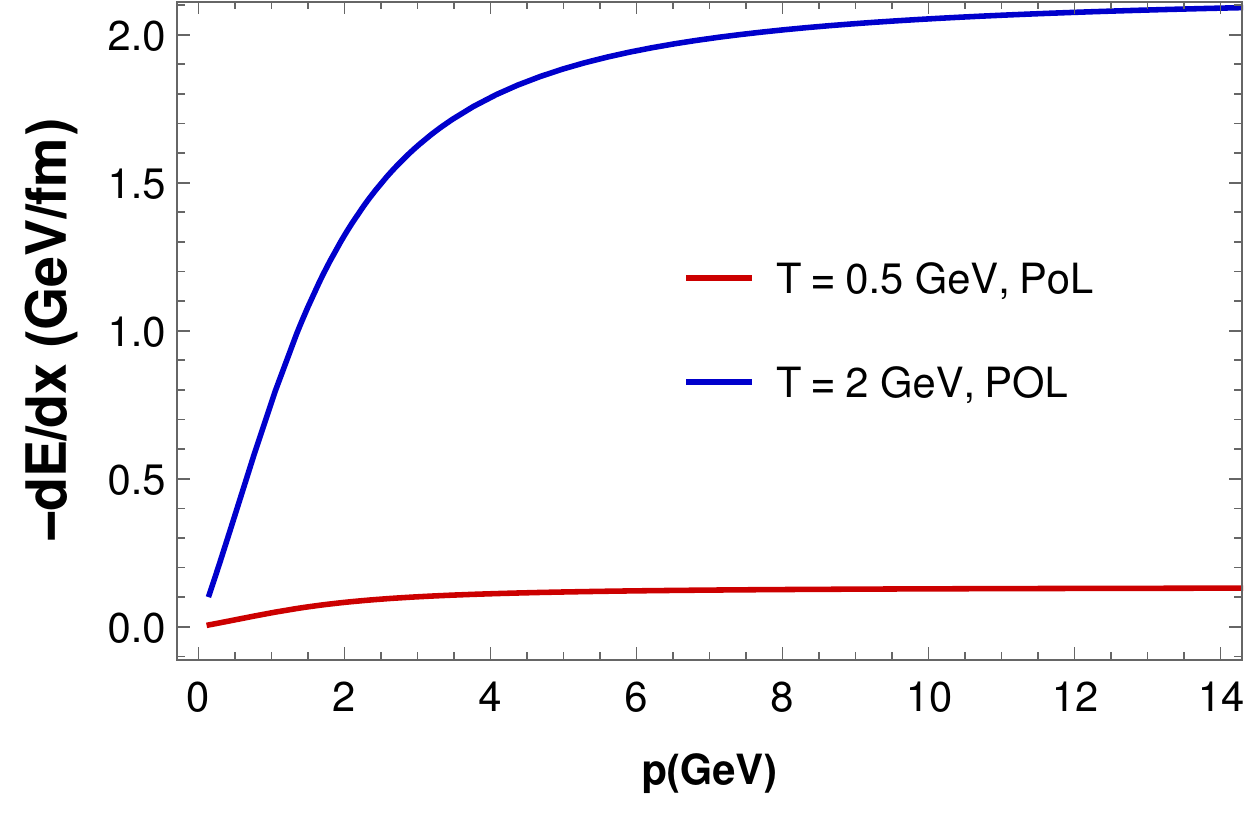}\\
	\includegraphics[height=5cm,width=6.0cm]{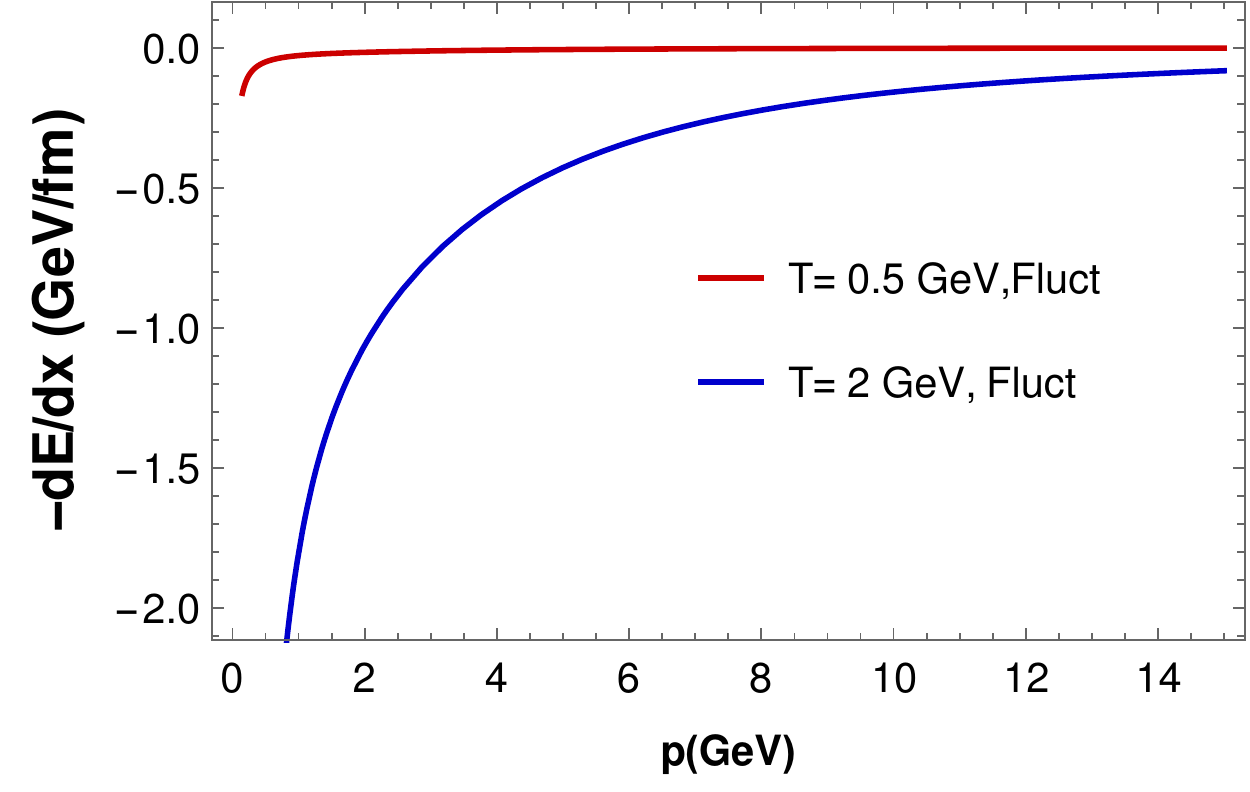}\\
	\includegraphics[height=5cm,width=6.0cm]{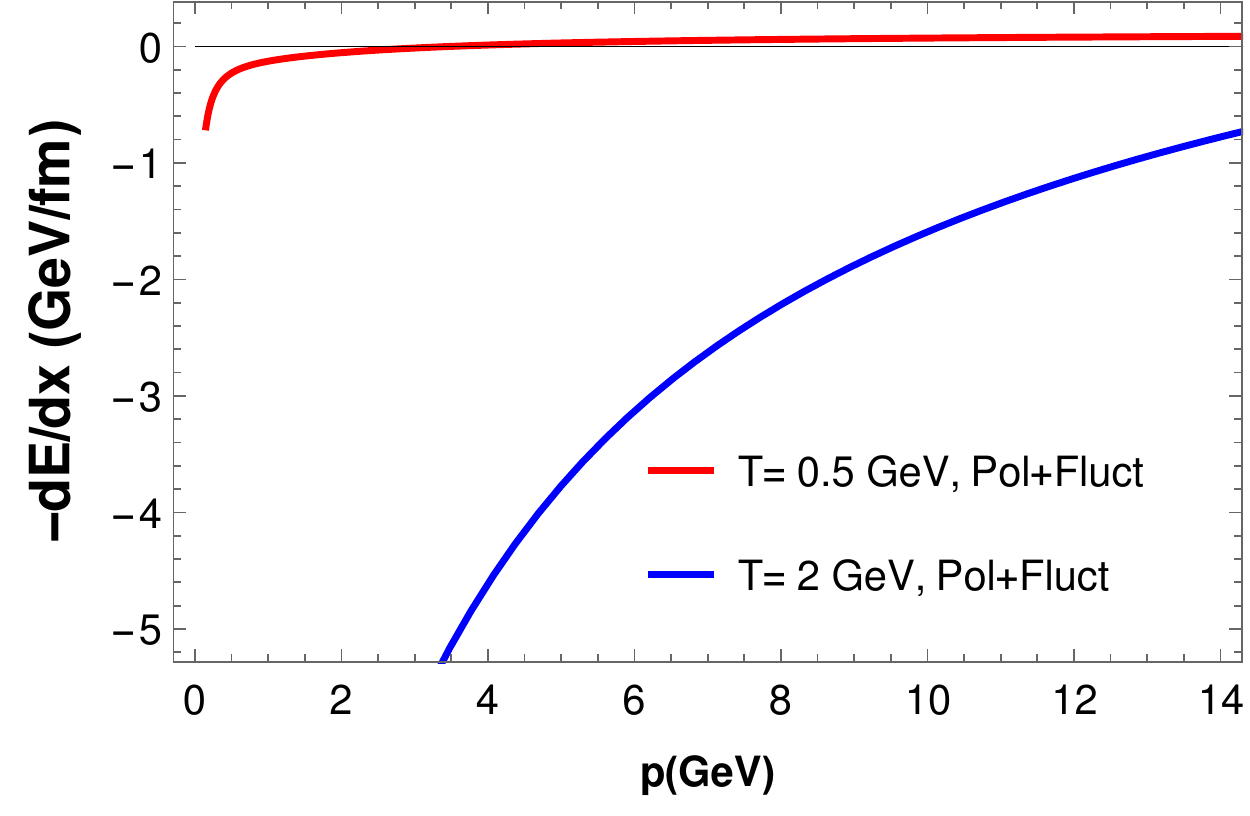}
	\caption{Energy change of charm quark versus the initial momentum, at $T=0.5$ GeV (red lines) and $T=2$ GeV (blue lines).
Upper and middle panels correspond to the polarization and fluctuations contributions respectively,
while the lower panel corresponds to the sum of the two contributions.	
	Results correspond to $\KMAX=2m_D$.\label{Fig:versp}}
\end{figure}

In Fig.~\ref{Fig:versp} we plot the energy change due to polarization (upper panel) and fluctuations (middle panel)
of charm quarks versus the initial momentum, for two representative values of temperature and for $\KMAX=2m_D$
(results for $\KMAX=m_D$ are similar to those shown here).
At relatively low temperature the energy loss dominates over energy gain for $p\gtrsim 2$ GeV, 
while for higher temperatures the energy gain due to the interaction with the fluctuating gluon field is more important
than the energy loss.

\subsection{Beauty}

\begin{figure} [t!]
	\centering
	\includegraphics[height=6cm,width=7.0cm]{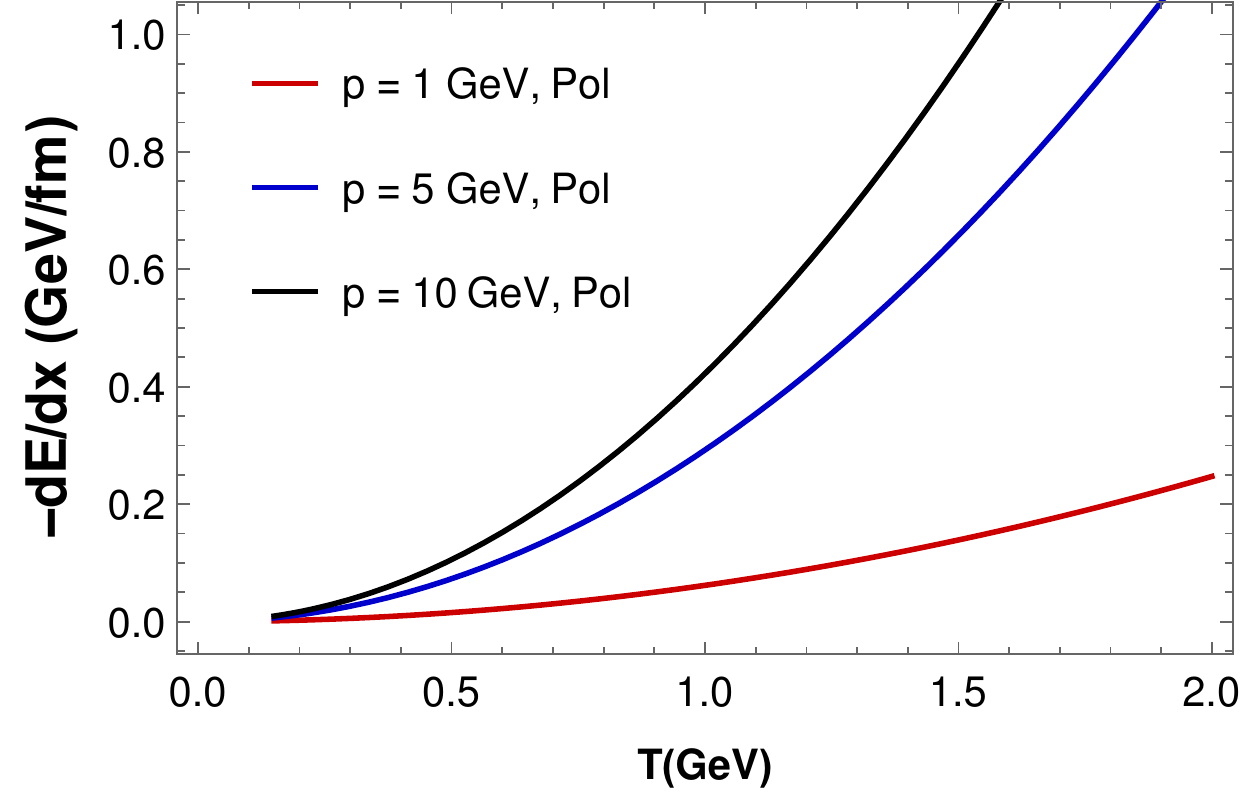}\\
	\includegraphics[height=6cm,width=7.0cm]{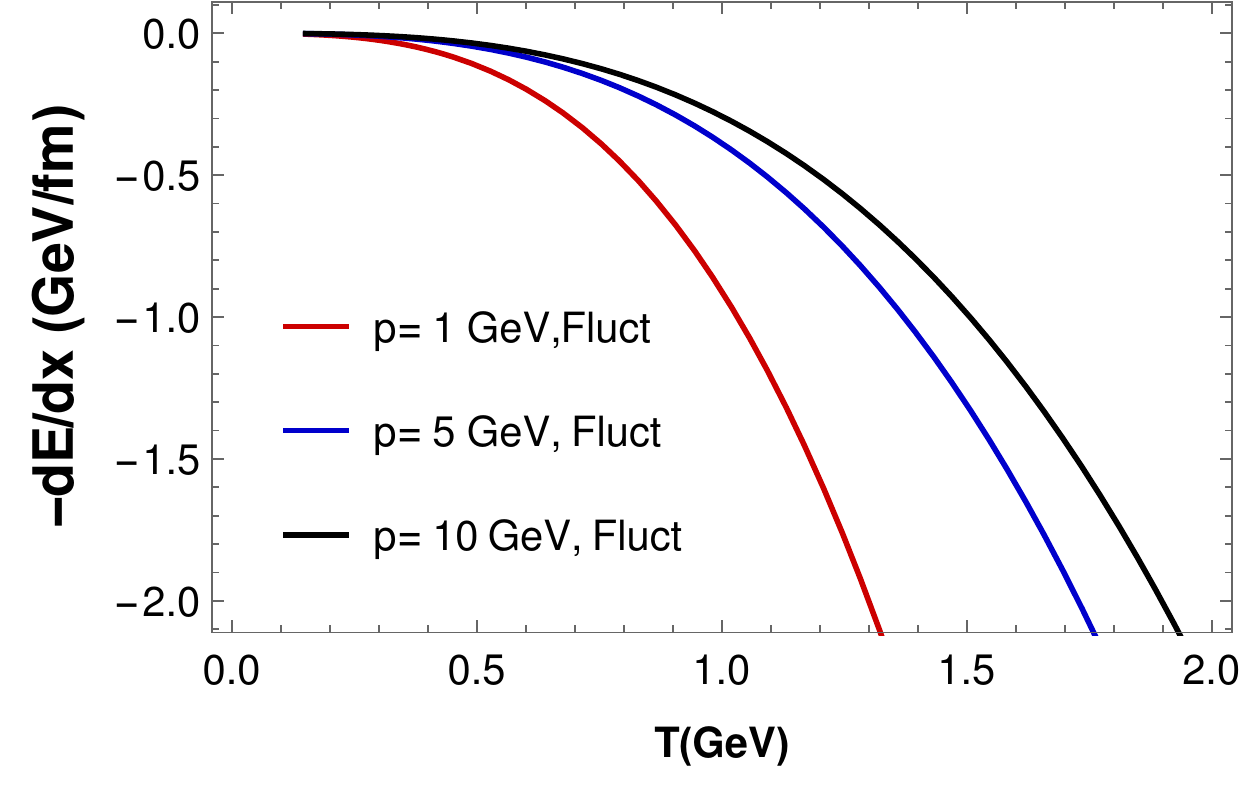}\\
	\includegraphics[height=6cm,width=7.0cm]{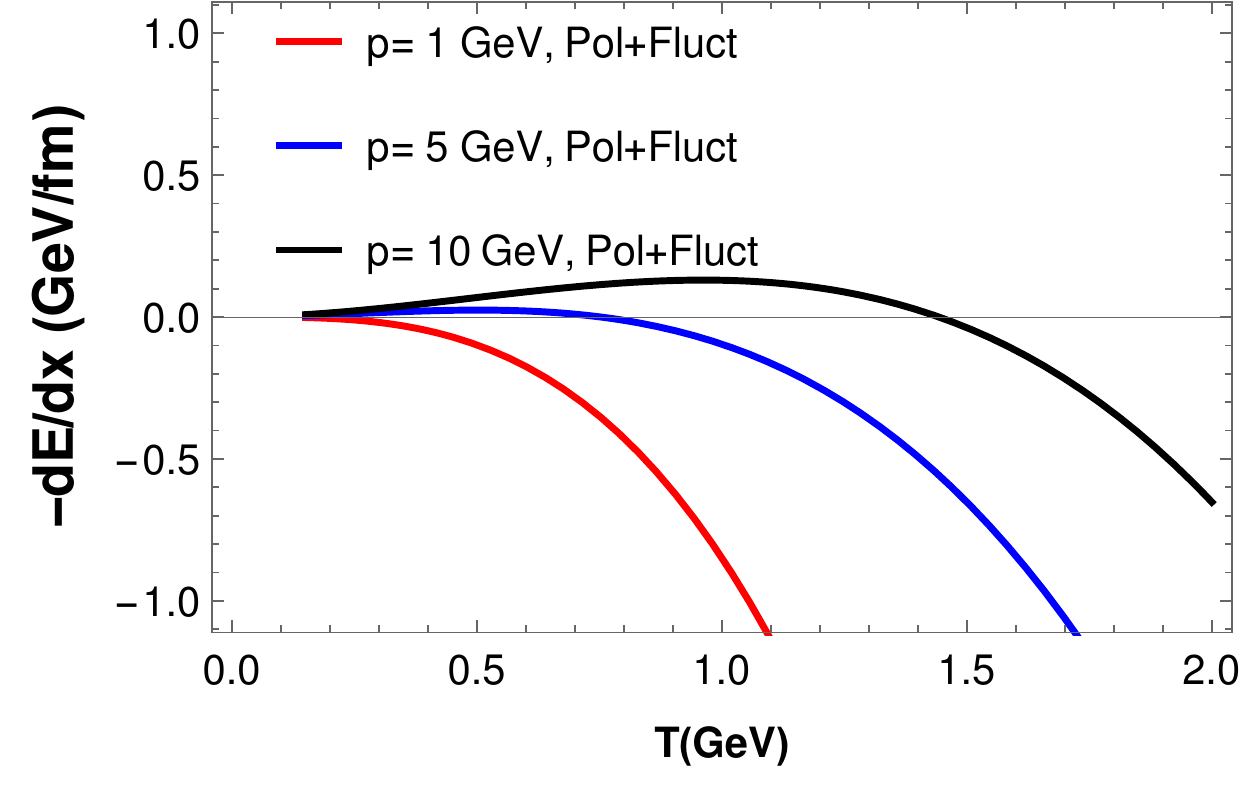}
	\caption{Energy change beauty due to polarization (upper panel)
	and fluctuations (middle panel), as well as the combination of the two (lower panel). 
	Results correspond to $\KMAX=2m_D$.\label{fig:full_beauty}}
\end{figure}

In this subsection we report on the analysis of energy loss and gain of beauty quarks in the hot medium;
since the qualitative picture is unchanged with respect to that of the charm quark, here we limit ourselves to
present only a few representative results.
In Fig.~\ref{fig:full_beauty} we plot energy change induced by polarization (upper panel),
interaction with fluctuating medium (middle panel) and total (lower panel) versus temperature,
for three values of the initial beauty quark momentum;
the results correspond to $\KMAX=2m_D$.
Clearly, there is some quantitative difference between charm and beauty, due to the different masses of the two quarks, {\it e.g.,}
for the given values of parameters, beauty quark loses less energy in the case of polarization and also gains less energy in the case of fluctuation as compared to charm quark.
Overall,	the combined effect of polarization and fluctuations on beauty results in an energy gain for $p/T\lesssim 1$
while energy loss becomes more important in the kinematic regime $p/T\gtrsim 1$.

\section{Conclusions}
\label{el:saf}
We have studied, within linear response theory, the energy change of heavy quarks in a hot thermalized QCD medium,
analyzing the 
combined effect of energy loss due to the polarization of the medium,
and energy gain due to interaction with the thermal fluctuations of the gluon field of the medium..
We have considered the effects on both charm and beauty quarks. 
This study has been inspired by a series of works on the propagation of heavy probes in the early stage of the
high energy nuclear collisions, in which the energy gain due to the diffusion in the evolving Glasma
is crucial to bend the initial pQCD spectrum of the heavy quarks before the formation of the 
quark-gluon plasma~\cite{Ruggieri:2018rzi,Liu:2019lac}.
Although we do not consider the Glasma in the present study, we think that the results found here support at least
qualitatively the diffusion-dominated scenario found in~\cite{Ruggieri:2018rzi,Liu:2019lac}: in fact, 
despite the fact that the evolving Glasma is a system out of thermal equilibrium, the diffusion of heavy color probes (see also Ref.~\cite{Boguslavski:2020tqz}) in it
is not very different from the diffusion in a Brownian motion, at least when an average over the full heavy quark spectrum is taken:
because of this similarity, it is likely that the results on diffusion in a fluctuating medium studied here can be applied
qualitatively to the diffusion in the evolving Glasma as well.

We have found that in the kinematic regime $p/T\lesssim 1$, where $p$ is the initial heavy quark momentum
and $T$ the temperature of the medium, energy gain dominates of the energy loss, and the situation inverts in the
complementary regime $p/T\gtrsim 1$.
These results are consistent with previous literature~\cite{Jamal:2019svc,Chakraborty:2006db}.
If we applied these conclusions to the early stages of high energy nuclear collisions,
our findings would suggest a diffusion dominated propagation for $p\lesssim 2$ GeV
while energy loss would be substantial for $p\gtrsim 10$ GeV, while in between
there would be a balance between the two.

The results may have a significant impact on the experimental observables like the nuclear suppression factor 
and elliptic flow~\cite{Rapp:2018qla,Dong:2019unq}  of heavy mesons produced at RHIC and LHC energies both for the nucleus-nucleus and p-nucleus collisions. 
Also a thorough understanding of the initial stage dynamics is a timely fundamental task and may affect 
observables like the  triggered $D- \bar D$ angular correlation~\cite{Nahrgang:2013saa} and the heavy quark directed flow $v_1$~\cite{Das:2016cwd}.
Apart from this, as it is well known that the thermal systems have comparatively weaker fluctuations than the 
non equilibrated systems. Therefore, incorporating the momentum anisotropy (which remains inevitible throughout 
the medium evolution) and also viscosity while modelling the medium~\cite{Chandra:2015gma,Das:2012ck} in the current study 
will bring us much closer to the real picture of the high energy nuclear collision. Hence, it will be an immidiate 
future extension to the current work.

\begin{acknowledgments}
	
M. R.  acknowledges John Petrucci for inspiration.
The work of M. R. and S. K. D. are supported by the National Science Foundation of China (Grants No.11805087 and No. 11875153)
and by the Fundamental Research Funds for the Central Universities (grant number 862946).
\end{acknowledgments}

\end{document}